\documentclass[conference]{IEEEtran}

\usepackage{graphicx} 
\usepackage{tcolorbox} 
\usepackage{tabularx} 
\usepackage{biblatex}
\usepackage{multirow} 
\usepackage{booktabs}
\usepackage{amssymb}
\usepackage{pifont}
\newcommand{\cmark}{\ding{51}}%
\usepackage{tikz}
\usepackage{listings}
\usepackage{colortbl}
\usepackage{xcolor}
\usepackage{url}
\usepackage{breakurl}
\usepackage{acronym}
\usepackage[scaled=.8]{beramono}
\usepackage{xspace}
\usepackage[T1]{fontenc}  
\usepackage{cleveref}  
\usepackage{paralist}  
\usepackage{array}  
\usepackage{diagbox}

\newacro{sbom}[SBOM]{Software Bill of Materials}

\newcommand{\ssc}{Software Supply Chain\xspace}
\newcommand{\sbom}{\ac{sbom}\xspace}
\newcommand{\sboms}{\acp{sbom}\xspace}
\newcommand{\cdx}{CycloneDX\xspace}
\newcommand{\spdx}{SPDX\xspace}
\newcommand{\tool}{\sbom generation tool\xspace}
\newcommand{\tools}{\sbom generation tools\xspace}
\newcommand{\python}{Python\xspace}
\newcommand{\pythontools}{package managers\xspace}
\newcommand{\Pythontools}{Package managers\xspace}
\newcommand{\pythontool}{package manager\xspace}

\newcommand{\unversioned}{unversioned\xspace}

\newcommand{\lock}{lockfile\xspace}
\newcommand{\locks}{lockfiles\xspace}
\newcommand{\req}{\texttt{requirements.txt}\xspace}
\newcommand{\pyproject}{\texttt{pyproject.toml}\xspace}
\newcommand{\setup}{\texttt{setup.py}\xspace}

\newcommand{\syft}{Syft\xspace}
\newcommand{\trivy}{Trivy\xspace}
\newcommand{\cdxgen}{Cdxgen\xspace}
\newcommand{\ort}{Ort\xspace}

\newcommand{\syftcolor}{\tikz\draw[magenta,fill=magenta] (0,0) circle (.9ex);\xspace}
\newcommand{\trivycolor}{\tikz\draw[blue,fill=blue] (0,0) rectangle (1.8ex, 1.8ex);\xspace}
\newcommand{\cdxgencolor}{\tikz\draw[orange,fill=orange] (0.2,0) -- (0.4,0) -- (0.3,0.2) -- cycle;\xspace}
\newcommand{\ortcolor}{\tikz\draw[green,fill=green] (0.2,0) -- (0.4,0) -- (0.2,0.2) -- (0.4,0.2) -- cycle;\xspace}

\newcommand{\hatch}{Hatch\xspace}
\newcommand{\pdm}{Pdm\xspace}

\newcommand{\poetry}{Poetry\xspace}
\newcommand{\pipenv}{Pipenv\xspace}

\newcommand{\backend}{back-end\xspace}

\newcommand{\frontend}{front-end\xspace}

\newcommand{\backends}{back-ends\xspace}
\newcommand{\Backends}{Back-ends\xspace}
\newcommand{\frontends}{front-ends\xspace}

\newcommand{\y}{\cmark}
\newcommand{\na}{NA\xspace}

\newcommand{\vclause}[1]{\texttt{#1}}
\newcommand{\veq}{\vclause{==}}
\newcommand{\vcomp}{\vclause{\textasciitilde=}}
\newcommand{\vle}{\vclause{\textless=}}
\newcommand{\vge}{\vclause{>=}}
\newcommand{\vlt}{\vclause{\textless}}
\newcommand{\vgt}{\vclause{>}}
\newcommand{\vneq}{\vclause{!=}}

\newcommand{\pheader}[2]{\multicolumn{1}{p{#1}}{\textbf{#2}}}  

\title{SBOM Generation Tools in the Python Ecosystem:\\an In-Detail Analysis}
\author{\IEEEauthorblockN{Serena Cofano}
\IEEEauthorblockA{\textit{IMT School for Advanced Studies Lucca}; \\ 
\textit{University of Genoa}\\
Genoa, Italy \\
0009-0006-6539-9931 \\
serena.cofano@imtlucca.it}
\and
\IEEEauthorblockN{Giacomo Benedetti}
\IEEEauthorblockA{
\textit{University of Genoa}\\
Genoa, Italy \\
0000-0003-2609-6787 \\
giacomo.benedetti@dibris.unige.it}
\and
\IEEEauthorblockN{Matteo Dell'Amico}
\IEEEauthorblockA{\textit{University of Genoa} \\
Genoa, Italy \\
0000-0003-3152-4993 \\
matteo.dellamico@unige.it}
}

\date{January 2024}

\begin{document}

\maketitle

\begin{abstract}
    Software Bills of Material (SBOMs), which improve transparency by listing the components constituting software, are a key countermeasure to the mounting problem of \ssc attacks.
    SBOM generation tools take project source files and provide an SBOM as output,
    interacting with the software ecosystem.
    
    While SBOMs are a substantial improvement for security practitioners, providing a complete and correct SBOM is still an open problem.
    This paper investigates the causes of the issues affecting SBOM completeness and correctness, focusing on the PyPI ecosystem.
    We analyze four popular SBOM generation tools using the CycloneDX standard. 
    
    Our analysis highlights issues related to dependency versions, metadata files, remote dependencies, and optional dependencies.
    Additionally, we identified a systematic issue with the lack of standards for metadata in the PyPI ecosystem.
    This includes inconsistencies in the presence of metadata files as well as variations in how their content is formatted.
\end{abstract}

\acresetall  

\section{Introduction}

Recent \ssc{} attacks~\cite{EnisaThreatSSC2021}, such as the \emph{SolarWinds} incident~\cite{martinez2021software} and the \emph{xz} attack~\cite{xz-attack}, have drawn significant interest from academics, industry players, and governmental organizations.
The requirement for increased transparency on the \ssc{}, (i.e., the visibility on its components~\cite{okafor2022secureDesignProperties}), has given rise to the \ac{sbom}~\cite{CisaSBOMDef} as a defense against these attacks.

An \sbom{} is a comprehensive list of the components constituting a piece of software~\cite{ntiaSBOMdef}.
Maintainers can use this list to quickly identify malicious or vulnerable software packages by feeding it into security analysis tools to find vulnerabilities.
The US Government identified the \ac{sbom} as a primary tool for transparency and made it compulsory for government software in 2021~\cite{EOnist2021}. On March 12th, 2024, the European Parliament approved the EU Cyber Resilience Act (CRA)~\cite{cra}.
The CRA uses the \ac{sbom} to describe, record, and monitor the security of products. 

Numerous studies have been conducted to examine the practitioners' perceptions of the \ac{sbom}~\cite{StalnakerAway2024, XiaEmpiricalStudySBOM2023, Zahan2023}.
Most users who include the \sbom into their security analyses agree that the presence of false positives and negatives makes \sbom quality a recurrent problem. 
An \sbom{}, to be useful from the point of view of security, should be \textit{complete}, i.e., include all the components, and \textit{correct}, i.e., provide exact information about them.
Tools that generate the \sbom are often responsible for losing these properties.
Although the issue is well known, few measures have been implemented, and little research has been done on the study of generation methodologies.
Some works propose a study of the performance of the most commonly used open-source generation tools. 
They mostly focus on a single language, such as Java~\cite{Balliu2023} and Javascript~\cite{Rabbi2024}. 
The decision to focus on a single language is driven by the fact that \sbom generation is significantly influenced by how the ecosystem manages a project's dependencies.
Other works aim to give an overview of \ac{sbom} generation: Yu et al.~\cite{metadataSbomGen2024} perform a large-scale differential analysis on four \sbom generation tools. The differential analysis allows them not to have to deal with ground truth and thus be able to consider multiple languages. On the other hand, such an approach has less detail about each individual language and the causes that lead to different tools producing different \sboms for the same piece of software.

Our study focuses on the elements that primarily affect \sbom creation in \python projects in terms of correctness and completeness.
In this work, we identify problems in \ac{sbom} generation and determine their root causes in both ecosystem and \tools.

The choice of \python is driven by the following reasons: 
\begin{enumerate}[\it (i)]
    \item Its widespread use \cite{pythonUsageStatistic, state-of-opensource-octoverse, decan_topology_2016}.
    \item Its flexibility: there is no standard for project creation, only guidelines (PEPs) that may not be followed. There are several tools for creating and managing \python projects and dependencies, and these tools require metadata files that specify dependencies in different ways;
    this flexibility has prompted the community to develop package managers capable of creating and managing \python project and its dependencies. However, these tools can create projects with differing metadata files containing project dependencies.
    \item Dynamic dependency resolution. Developers can specify a range of versions or no version at all for dependencies, and this will be resolved at installation time. For example, when the version is not specified, the \pythontool will use the last available version. 
    In this case, the version will be known only after the installation and depends on when the package is installed.
\end{enumerate}

In detail, we provide the following contribution:
\begin{itemize}
    \item A study on the impact of the \python ecosystem on the generation of \sboms. In detail, we investigate how, given the same set of dependencies, the methods used to generate the \python project influence the \sbom computed on the projects. For this goal, we create a set of \python projects with the same dependencies but using different \pythontools. 
    \item A study on how the approach used by \tools changes the final output of the \sbom. We select four open-source \tools and we run them on the projects. 
\end{itemize}

We find substantial differences in the generated \sboms, due to tool-related causes, different Python project configurations, or intrinsic to the ecosystem.
We identify the lack of standards in the \python ecosystem and the defects in \tools as main takeaways. 
In particular, we recommend that 
\begin{inparaenum}[\it (i)]
    \item Python package managers provide metadata in a consistent and stable format, and
    \item \tools improve the support for the most recent and recommended project configuration file, fixing the issues we identify. 
\end{inparaenum}

\section{Background}
\label{sec:background}
The security of a \ssc depends on the security of its components. To monitor each component and conduct security analysis, it's essential to have a complete and correct \ac{sbom}. 
The \ac{sbom} serves as input for security analysis tools, allowing for the identification of vulnerabilities~\cite{anchoreSBOMUse}. 
This is crucial for security maintainers to promptly address any issues.
This section contains notions of what a \ac{sbom} is and how it can be generated. Also, since our study focuses on \sboms related to \python projects, it describes how \python projects are composed and how \pythontools handle dependencies and resolve versions.

\subsection{Software Bill Of Material}
\label{sec:background-sbom}
The \ac{sbom} is an inventory of software components and dependencies, information about those components, and their hierarchical relationships \cite{ntiaSBOMdef}.
There are multiple standards for \ac{sbom}.
The most used are \cdx~\cite{CyclondeDx} and \spdx~\cite{SPDX}.
We focus our work on \cdx,
because it is designed primarily to generate the \ac{sbom} and is more suited for providing precise software components, which is why it is more effective for vulnerability management. Conversely, \spdx was designed as a tool for managing licenses, and it is now primarily used for development~\cite{CdxVsSpdx}.

The \cdx standard supports JSON, XML, and Protobuf formats as output~\cite{cyclonedxSpec}.
The standard covers multiple aspects of the \ssc composition, such as the human factor, licensing, vulnerabilities, metadata, and software components.
In particular, the latter is about the software dependencies represented by the \ac{sbom}.
A \textit{component} is represented along with information useful to identify it, e.g., name, version, package URL (purl), and licenses.
A purl is a URL string used to identify and locate a software package in an universal and uniform way across programming languages, package managers, packaging conventions, tools, APIs, and databases~\cite{purl-spec}.
\acp{sbom} are produced by \ac{sbom} generators, which are either static or dynamic.
Static tools generate \acp{sbom} by examining the binary or package dependencies without executing the software.
They aim to efficiently scan large codebases, extract metadata, and identify a wide range of components and licenses, with a low impact on computing resources.
Dynamic tools generate \acp{sbom} by simulating and interacting with the system where the software is installed.
The approaches used by these tools include environment scanning, i.e., simulating an installation by creating a virtual environment and looking for installed dependencies, and runtime monitoring, i.e., instrumenting the execution.
While computationally more expensive, dynamic \ac{sbom} generation should provide a more accurate reflection of the software’s actual composition in production environments.
However, this technique is not based on a real installation process, but on a simulated one which may happen at a different time from real installations.
As a result, there may be differences in the installed dependencies based on the particularities of the installation environment's setup or on installation time.

\subsection{Python}
\label{sec:background-python}
PyPI\footnote{\url{https://pypi.org/}} is the \python ecosystem's package registry.
It indexes \python packages, allowing developers to add them as dependencies to their applications.
\python packages can be distributed as either build or source distributions, which are both archives: build (also known as \emph{wheel}) distributions are zip archives, whereas source distributions are tarballs (i.e., archives compressed with the \texttt{tar} utility).
A \python project is composed of source code and metadata files.
Metadata files are required to define the project properties and to enable the package manager to build the project.
These files can be \pyproject, \setup, \req, and \locks.
The \pyproject and \setup files are mandatory for a project to be defined as a package and uploaded to the PyPI registry.
The former is the standard metadata file since 2016~\cite{PEP518}, while the latter is a legacy version still in use.
The \req file contains a list of the direct dependencies necessary for the correct functioning of the final project.
Finally, the \emph{\lock} is the most complete representation of the software.
It contains dependencies with exact versions, including transitive dependencies (i.e., dependencies of dependencies), and guarantees integrity by hashing package content.
It is automatically generated while dependencies are declared in the project.

The metadata files include dependencies specified with versions in various formats~\cite{PEP440}. Versions can be either \emph{pinned} or \emph{unpinned}. Pinned versions specify the exact version using the version matching operator ``\veq'' and are the way dependencies are represented in lockfiles. Alternatively, versions can be pinned or left unpinned in the \req, \pyproject, and \setup files.
The \textit{unpinned} dependencies are either \emph{unversioned} or \emph{constrained}, i.e., they specify a range of possible versions.

Constrained dependencies can have different clauses: the compatible release clause ``\vcomp'',\footnote{\url{https://packaging.python.org/en/latest/specifications/version-specifiers/\#compatible-release}} which matches any candidate version that is expected to be compatible with the specified version, the version exclusion clause ``\vneq'', and the inclusive and exclusive comparison clauses ``\vle'', ``\vge'', ``\vlt'', and ``\vgt''.

The metadata files provide a way to distinguish \emph{required} dependencies from \emph{optional} ones.
Required dependencies are those necessary to run the project with the core functionalities.
Optional dependencies are not fundamental to the correct functioning of the project but are either used during the development phase (\emph{development} dependencies), or are related to additional features that can be made available to the user.
The final user of a Python project can decide whether to install any optional dependency.

While the \pyproject and the \lock group the two in different sections of the files, \req does not distinguish the two. In this case, the alternative is to list optional dependencies in a separate file.

Different \pythontools can handle Python projects. \Pythontools can be divided in \emph{\frontends} and \emph{\backends}.
The \frontend manages metadata and dependencies and communicates with the \backend using the project's metadata file (\pyproject or \setup). The \backend is responsible for building the actual project in a distributable package. 
It's important to choose the \backend that aligns with the selected \frontend, as not every \frontend supports every \backend. Additionally, front- and back-ends can differ in their support for non-Python code, the presence of a \lock, and their ability to obtain packages from a registry other than PyPI.

\section{Study methodology}
\label{sec:methodology}
In this section we illustrate how we conduct our study on \sbom completeness and correctness in \python projects.
Our goal is to identify the issues with the generation of the \ac{sbom} and determine their relation with the \tool or with the ecosystem.

We follow a three-step methodology.
The first step concerns the \emph{experimental setup}, in which we create a dataset of \python projects and select a set of \tools for running the experiment.
The second step is \emph{\sbom generation} for our \python projects dataset using the selected tools.
The third step is \emph{analysis};
this phase involves inspecting and comparing the \sboms to discover significant issues in their generation and their causes.
This is done by thoroughly diving into the implementation of the selected \tools and comparing it with the specific \python projects for which the tool is executed.

\subsection{Experimental Setup}
\label{sec:experimental-setup}
\begin{table}[t]
    \centering
    \caption{\Backends (\textsc{Be}) and \frontends (\textsc{Fe}) selected for python projects generation}
    \begin{tabular}{cccccc}
        \toprule
        \diagbox{\textsc{Be}}{\textsc{Fe}} & \textbf{Pip} & \textbf{Hatch} & \textbf{Pdm} & \textbf{Pipenv} & \textbf{Poetry} \\
        \midrule
        \textbf{Hatchling}  & \y  & \y & \y & - & - \\
        \textbf{Flit}  & -  & - & \y & -  & - \\
        \textbf{Pdm} & \y & \y & \y & \y & -\\
        \textbf{Poetry}  & -  & - & - & - & \y \\
        \textbf{Setuptools}  & \y  & \y & \y & - & -\\
        \bottomrule
    \end{tabular}
    \label{table:python_projects}
\end{table}

This step consists of creating a dataset\footnote{The dataset will be publicly available upon publication.} of \python projects and selecting \tools{}.
\subsubsection{Dataset creation}
Our \python projects dataset is synthetic.
We make this choice rather than using existing active projects because we:
\begin{enumerate}[\it (i)]
    \item want to test patterns of dependency handling in the PyPI ecosystem by observing \tools on projects with the same dependencies but created with different technologies;
    \item do not have a way to create a trusted ground truth for the components of a large existing project~\cite{metadataSbomGen2024}.
\end{enumerate}

Therefore, for the creation of the dataset we have to \begin{inparaenum}[\it (i)] 
    \item select the tools in \python that allow us to create a project, and
    \item select the content of the projects, i.e., what dependencies they should contain.
\end{inparaenum}

We base the project setups on an assessment of online articles on \pythontools to match reality.
The following factors are taken into account while choosing the \frontend and \backend:
\begin{itemize}
    \item number of \python projects that adopt them;
    \item number of articles that use them in a comparison or suggest them for building a \python package;
    \item amount of documentation present about them;
    \item activity on maintenance of them: we take into account only actively maintained \pythontools;
    \item scope of the \pythontools: we consider those limited to the managing \python projects.
\end{itemize}

Due to these criteria, we select \emph{Setuptools} and \emph{Poetry}, which are the most used tools~\cite{pyOpenSci}.
Moreover, we select \emph{Pdm}, \emph{Flit}, \emph{Pipenv}, and \emph{Hatch} which are less used, but are cited in articles about \pythontools for \python~\cite{packagingTutorial, caironiComparison, toolRecommendationPyPA, fuzzyComparison}.
We end up with 5 \backends and 5 \frontends. Of the 25 possible combinations, 12 are compatible, as shown in \Cref{table:python_projects}. We analyze all of them. 

Next, we select the dependencies to add to each project.
The purpose is to study \tools behavior in as many case scenarios as possible.
As a result, we choose to include pinned, \unversioned, and unconstrained.
In addition, we consider origins other than the default PyPI, such as GitHub and other registries identified with URLs.
We also consider optional and development dependencies.
This allows us to choose the dependencies from \Cref{table:dependencies}.
Only one dependency, Numpy, is imported and used in the code. This helps us understand whether there is a difference between just declaring a dependency and actually using it.

Each element in the sample is created following this procedure:
\begin{enumerate}
    \item Initialize the project according to the selected \frontend.
    \item Add the selected dependencies.
    \item Add a main file with example code importing an installed dependency, and using it.
    \item Use the \pythontool to package the project in a distributable wheel and a source tarball (to test the correctness of the project setup).
\end{enumerate}

\begin{table}
    \centering
    \caption{Set of installed dependencies inside of each project in the sample.}
    \begin{tabular}{ccc}
        \toprule
        \textbf{Name} & \textbf{Version} & \textbf{Type} \\
        \midrule
        \textit{numpy} & Unversioned & Imported and used in the code \\
        \textit{docopt} & Pinned & Remote\\
        \textit{black} &  Pinned & Remote from Git\\
        \textit{seaborn} & Pinned & Not used in the code\\
        \textit{matplotlib} & Contrained  & Not used in the code\\
        \textit{urllib3} & Unversioned & Not used in the code\\
        \bottomrule
    \end{tabular}
    
    \label{table:dependencies}
\end{table}

\subsubsection{\tools selection}
We select \tools among those adopting the \cdx standard.
From the complete list of 219 tools hosted by the \cdx{} Tool Center,\footnote{\url{https://cyclonedx.org/tool-center/}}
we filter for those having:
\begin{itemize}
    \item an open-source implementation;
    \item a command line interface;
    \item a recent release (i.e., at least a new release in 2023);
    \item Python support;
    \item a standalone implementation (i.e., not being an extension of other tools);
    \item support for standard installation scenarios (e.g., we excluded tools scanning just container images);
    \item support for recent \cdx{} versions (at least version 1.3).
\end{itemize}
We select \emph{Trivy}, \emph{Syft}, \emph{Cdxgen}, and \emph{Ort}.

\trivy and \syft are static tools, which only read and parse a selection of metadata files containing the dependencies.
They differ in how they resolve versions; 
more details and the impact of this will be discussed in \Cref{results:version}.

\cdxgen has a hybrid approach between static and dynamic.
It simulates an installation by creating a \python virtual environment and installing the dependencies;
it then scans the content of the environment and fetches the installed dependencies to generate the \sbom. If this procedure fails, it proceeds parsing metadata files.

\ort parses only the \req file as metadata; if it is not present but the project was created using \poetry and \pipenv, it uses the capability of those tools to generate \req.
Since transitive dependencies are not present in \req, \ort exploits the PIP logic and directly queries PyPI to get them.

The main characteristics of the tools are summarized in Table~\ref{table:tools}.

\begin{table}[t]
    \centering
    \caption{List of selected \tools}
    \begin{tabular}{lrr>{\raggedright\arraybackslash}p{9em}}
        \toprule
        & \pheader{3.5em}{Tool version} & \pheader{3.5em}{CycloneDX version} & \textbf{Generation method} \\
        \midrule
        \textbf{Trivy} & 0.49.1 & 1.5  & metadata parsing \\
        \textbf{Syft} & 1.0.1 & 1.5 & metadata parsing \\
        \textbf{Cdxgen}  & 10.2.2 & 1.5  & metadata parsing, environment analysis \\
        \textbf{Ort}  & 17.0.0 & 1.5  & metadata parsing, PyPI querying \\
        \bottomrule
    \end{tabular}
    \label{table:tools}
\end{table}
\subsection{SBOM Generation}
\label{sec:sbom-generation}
We generate \sboms for each project in our sample.
\syft and \trivy run on the host machine, while \cdxgen and \ort run on a Docker container. 

We obtain a total of 48 \sboms.
The 12 \sboms produced by \ort were in XML format; we converted them to JSON using a \python script.
We manually verified that the conversion did not affect the \sbom content.

\subsection{Analysis}
\label{sec:analysis}

This phase is divided into two sub-phases: 
\begin{inparaenum}[\it (i)]
\item observing the \sboms and identifying the correctness (i.e., wrong versions) and completeness (i.e., lost dependencies) issues;
\item looking for the causes of these issues, which are either due to the ecosystem itself (i.e., to how the projects are structured), or to the \sbom generation tools (because of bugs and/or root causes of the methodology used).
\end{inparaenum}

\subsubsection{\sbom Analysis} 
We manually analyze each \ac{sbom}. We verify completeness by checking whether direct, transitive, remote, and optional dependencies are present;
for correctness, we check if the versions are exact.
To perform this task, since for unpinned dependencies we cannot determine a correct version, we compare the version numbers output by different \sboms. 

\subsubsection{Investigation of Causes}
After determining the critical issues in the generated \sboms, we identify the causes.
To do this we analyze tools through static code analysis, documentation study, and an assessment of the community reaction to some of the tool issues by searching for issues on GitHub.
From this analysis, we can determine whether an issue is due to implementation or methodological flaws of the tools, or from the way the \pythontools handle dependencies.
We discuss the results in \Cref{sec:results}.

\section{Results}
\label{sec:results}
\begin{table*}
    \centering
    \caption{Results overview.\\
        Notes: * only dependencies that are also used in the application code , ** it does not work due to implementation errors.\\
        Cdxgen = \protect\cdxgencolor , ort = \protect\ortcolor , syft = \protect\syftcolor , trivy = \protect\trivycolor}
    \begin{tabular}{cccccc}
    \toprule
        &\textbf{Hatch}& \textbf{Pdm} & \textbf{Pipenv} &  \textbf{Poetry} & \textbf{Pip}\\
        \midrule
        Find direct dependencies & \cdxgencolor* & \cdxgencolor & \cdxgencolor \syftcolor \trivycolor & \cdxgencolor \ortcolor \syftcolor \trivycolor & \cdxgencolor \ortcolor \syftcolor \trivycolor \\
        Find transitive dependencies & - & \cdxgencolor & \cdxgencolor \syftcolor \trivycolor & \cdxgencolor \ortcolor \syftcolor \trivycolor & \cdxgencolor \ortcolor \syftcolor \trivycolor \\
        Find remote dependencies & - & \cdxgencolor & \cdxgencolor \syftcolor \trivycolor & \cdxgencolor \syftcolor \trivycolor & \cdxgencolor \\
        Find optional dependencies & - & \cdxgencolor & \syftcolor \trivycolor & \cdxgencolor \ortcolor \syftcolor \trivycolor & \cdxgencolor \ortcolor \syftcolor \trivycolor \\
        Implement resolution of not present versions  & \na & \na & \na & \na & \cdxgencolor \ortcolor\\
        Implement resolution of constrained versions & \na & \na & \na & \na & \cdxgencolor \ortcolor \syftcolor \\
        Implement parsing of lockfiles & \na & \cdxgencolor & \cdxgencolor \ortcolor** \syftcolor \trivycolor & \cdxgencolor \ortcolor \syftcolor \trivycolor & \na \\
        Implement parsing of requirements.txt & \na & \na & \na & \na & \cdxgencolor \ortcolor \syftcolor \trivycolor \\
        Implement parsing of pyproject.toml & - & - & - & \ortcolor \trivycolor & - \\
        \bottomrule
    \end{tabular}    
    \label{table:overview}
\end{table*}

In this section, we analyze \acp{sbom} for completeness and correctness, as discussed in Section~\ref{sec:methodology}. We present identified issues and their causes, differentiating between ecosystem and \tools origins. Table~\ref{table:overview} gives an overview of our findings, showing how \tools{} behave based on the \pythontool.

We identify two scenarios when these issues arise: \textit{version management} and \textit{metadata file handling}. 
Therefore, we group the issues into these two categories and consequently divide the section.
Table~\ref{tab:overall-issues} illustrates how issues within these categories impact the completeness and correctness of \sboms across the ecosystem (E) and the \tools{} (T).

Recall \Cref{sec:methodology}, we imported a dependency in the code, i.e., Numpy.
It is noteworthy that the source code is never analyzed by \tools, with the only exception of \cdxgen for the project generated by Hatch.
This is a fallback approach \cdxgen uses when there are not \emph{supported} metadata files.
We conducted some manual experiments on this behavior noticing that it works only in specific cases.
Since \sboms are not affected by the actual importing of a dependency in the source code, we do not report any specific result for that.
    
\subsection{Version Management Issues}
\label{results:version}
Version management is the process of assigning a version to a package. If the version is explicit, the tools should detect it; otherwise, they should determine it.

As discussed in \Cref{sec:background-python}, \pythontools allow declaring dependencies without specifying an exact version (E9).
Specifically, they allow omitting the version or indicating a range of possible versions; the exact version is defined at installation time. 
\tools implement techniques to resolve the version.

The tools based solely on static analysis of metadata files, \syft and \trivy, ignore \unversioned dependencies and do not include them in the final \ac{sbom} (T3), resulting in a loss of information. 

When the version is indicated through a range, \syft applies guessing techniques (T6).
Specifically, the guessing technique identifies comparison operators, discarding those not including ``\texttt{=}'' and converting the remaining ones to ``\veq''. For versions like ``\texttt{1.1.*}'', the asterisk is replaced with a ``\texttt{0}''.
Considering that \citeauthor{metadataSbomGen2024} found that on average only about 46\% of the dependencies in the \verb|requirements.txt| files are pinned~\cite{metadataSbomGen2024}, this technique leads to a significant number of false positives in the \ac{sbom}, affecting its correctness.

Tools such as \cdxgen and \ort try to exploit the same version resolution mechanism used by \pythontools.
\cdxgen simulates the installation, while 
\ort contacts the PyPI API.
Constraints are resolved in the same way that they are during installation.
Theoretically, this technique allows replicating what is done at installation time; however, it is valid only if installation and \ac{sbom} generation happen at the same time.
For example, if an unspecified version is resolved as ``latest'' at a particular moment and the project is installed later, the two ``latest'' versions may not coincide.
This would result in a loss of correctness and completeness.

\subsection{Metadata Files Management Issues}
\label{results:metadata}

\begin{table*}[t]
\centering
\caption{Comprehensive overview of issues, depending on tools (T\#) and ecosystem (E\#).}
\begin{tabular}{@{}lll@{}}
\toprule
\textbf{Property} & \textbf{Effect}               & \textbf{Issue}  \\ \midrule
\multirow{13}{*}{\textbf{Completeness}} &\multirow{13}{*}{\textbf{Missing Dependencies}} & \textbf{T1}   SBOM generation tool does not consider pyproject.toml file.                                                        \\ 
                                 & & \textbf{T2}   SBOM generation tool does not consider lockfile.                                                                   \\ 
                                 & & \textbf{T3}   SBOM generation tool ignores dependencies without pinned version.                                               \\ 
                                 & & \textbf{T4}   SBOM generation tool fails in correct parsing of the optional dependency.                                          \\ 
                                 & & \textbf{T5}   SBOM generation tool does not properly parse the URL of the packages.                                                \\ 
                                 & & \textbf{E1}   Ecosystem has two build interfaces, setup.py and pyproject.toml.                                                   \\ 
                                 & & \textbf{E2}   The use of a lockfile is not mandatory.                                                                            \\ 
                                 & & \textbf{E3}   The ecosystem does not provide a standard file name for the requirements.txt file.                                 \\ 
                                 & & \textbf{E4}   Metadata files contain only direct dependencies.                                                               \\ 
                                 & & \textbf{E5}   Package manager does not create a lockfile.                                                               \\ 
                                 & & \textbf{E6}   Lack of a univocal standard for declaring optional dependencies.                                                   \\ 
                                 & & \textbf{E7}   Package managers do not explicitly declare version of the package.                                                   \\ 
                                 \midrule
\multirow{7}{*}{\textbf{Correctness}} & \multirow{3}{*}{\textbf{Wrong Dependencies}}   & \textbf{T4}   SBOM generation tool fails in parsing of the optional dependency.                                          \\ 
                                & & \textbf{T6}  SBOM generation tool guesses the dependency’s version.                                                          \\ 
                                 & & \textbf{T7}   SBOM generation tool does not report the origin of the packages.                                                     \\
                                 \cmidrule(l){2-3}
& \multirow{4}{*}{\textbf{Missing Versions}}  & \textbf{T5}   SBOM generation tool does not properly parse the URL of the packages.                                                \\ 
                                & & \textbf{E7}   Package managers do not explicitly declare version of the package.                                                   \\ 
                                & & \textbf{E8}   The format for the lockfile is not standardized.                                                                       \\ 
                                 & & \textbf{E9}   Version can be omitted in metadata files.                                                                              \\ 
                                 
                                 \bottomrule
\end{tabular}
\label{tab:overall-issues}
\end{table*}

Metadata files management includes identifying them, parsing them, or using them to install dependencies in a simulated environment.

We categorize the discussion based on the specific files involved: \lock, \pyproject, and \req.
Each paragraph highlights the challenges and inconsistencies associated with these files, and how they impact the completeness and correctness of the SBOM.
Finally, we highlight some specific aspects of addressing optional and remote dependencies in metadata file handling.

\subsubsection{Lockfiles}
Most \tools{} rely on the \lock to collect dependencies.
As discussed in Section~\ref{sec:background}, a \lock is supposed to be the most complete representation of a software, freezing both direct and transitive dependencies to specific instances. 
These files, although detailed, lack a standardized format in the \python ecosystem (E8), making their usage difficult. \python projects can use various formats for lockfiles, such as those from \poetry, \pipenv, and \pdm, resulting in tools developing ad hoc implementations for each (E8).
When the tool does not implement a specific \lock, it cannot parse it and thus ignores it, losing the dependencies it contains (T2).
This lack of standardization causes variability and inconsistencies in the generated \sboms, affecting both completeness and correctness.
Moreover, the lockfile is not mandatory for \python projects (E2).
Thus, some \pythontools (e.g., \hatch) do not generate this file at all (E5).

\ort does not directly parse the \lock, but instead converts them to \req format via the \pythontools' commands.
This applies solely to \poetry and \pipenv --- i.e., ``\texttt{poetry export}'' and ``\texttt{pipenv~requirements}''. However, \ac{sbom} generation does not work for \pipenv due to a known implementation issue.\footnote{https://github.com/aboutcode-org/python-inspector/issues/11}

\subsubsection{\pyproject}
\pyproject file is a standard for modern Python projects, hence \pythontools create it during project initialization.
However, \tools generally ignore this file (T1), except for \ort and \trivy, which consider the \pyproject file generated by \poetry. They parse the table containing dependencies, which is formatted with a syntax specific to \poetry. 
This \tools approach leads to significant gaps in dependency collection, thus affecting both the completeness and correctness of the \ac{sbom}.

An example of this issue is the \hatch project, where neither lockfiles nor \req files are present. Even though the \pyproject file contains the full list of direct dependencies, the \sbom is empty because tools do not parse this file, resulting in a total loss of completeness.

Another issue is the coexistence of the old standard \setup alongside \pyproject (E1). While the \python community strongly encourages updating to the new standard~\cite{PEP518, PEP621}, some tools still support only \setup, resulting in incomplete \sboms for new projects.

\subsubsection{\req}
All of the tools we analyze are able to retrieve direct dependencies from this file. However, if a developer chooses a different name for this file, these tools cannot detect it due to the absence of a standardized naming convention (E3). This inconsistency leads to a loss of completeness as the information in the \req file might be ignored.

Apart from \lock, metadata files contain only direct dependencies by default (E4).
Moreover, \req files typically include only direct dependencies (E4).
Because of that, tools relying on static parsing of metadata files miss out on transitive dependencies when there is no \lock present, thus affecting the completeness and correctness of the \ac{sbom}.

\textbf{Remote Dependencies.}
Remote dependencies are univocally identified by: name, version, and URL.
This information can be (1) wrapped by the URL or (2) listed by field, depending on the \lock format.
Considering these two cases:
\begin{inparaenum}[(1)]
\item When SBOM generation tools have to extract the remote dependency information out of the URL, often fail to correctly parse the URL (T5), leading to missing versions and incomplete dependencies.
\item Because there are no standards in the format of lockfiles (E8), the parsing methods of some tools may not comply with how the dependency is declared.
This can lead to missing or incorrect dependency information.
For example, Pipenv \lock omits versions for remote dependencies, causing \tools to miss the dependency (E7).
\end{inparaenum}

\textbf{Optional Dependencies.}
The lack of a univocal standard for declaring optional dependencies (E6) causes misidentification and loss of these dependencies in the \ac{sbom}. 
\tools sometimes fail to correctly parse optional dependencies (T4), resulting in missing dependencies. 
This issue is particularly evident in \pipenv, where \syft and \trivy fail to detect optional dependencies due to the formatting of the \pipenv \lock.
The \lock divides dependencies into two groups: ``default'' and ``develop'', with the latter containing the optional dependencies. 
\syft and \trivy's parsing implementations only consider the ``default'' group, excluding optional dependencies from the \ac{sbom}. 
This causes a lack of correctness in the generated \ac{sbom}.

\section{Related Works}
Despite the emergence of the \sbom technology and the well-known problems of its generation \cite{StalnakerAway2024, Zahan2023, XiaEmpiricalStudySBOM2023}, there is little literature on the precise definition of these problems and identification of the causes.
We can classify the works on \sbom generation into two groups: those considering different programming languages and those focusing on a single language.

Mirakhorli et al.~\cite{mirakhorli2024landscape} perform an empirical analysis of tools related to \sboms. They classify the open- and closed-source tools based on their role, then focus on the tools for \ac{sbom} generation and analyze five open-source tools testing them on a Java control project. 
Among the main takeaways they identify the lack of a reliable ground truth, inconsistency among \sboms generated by different \tools and a low accuracy in case of dependencies malpractices, e.g., hard-coding or dynamically loading dependencies. 

Yu et al.~\cite{metadataSbomGen2024} conduct a large-scale differential analysis of the correctness of four popular SBOM generators on 7,876 open-source projects written in Python, Ruby, PHP, Java, Swift, C\#, Rust, Golang and JavaScript. The differential analysis makes it possible to avoid generating a ground truth at the cost of not evaluating the precision of the tools. Considering different languages allows them to have an overview of the correctness of the \tools, but fewer details. Among the contributions, they focus on \python by demonstrating a possible attack; however, they do not discuss the differences among projects that use different \pythontools.
Two works focus only on a single language: Balliu et al.~\cite{Balliu2023} focus on Java, Rabbi et al.~\cite{Rabbi2024} focus on Javascript;
they are related, with the second being inspired by the first.
In this case, the authors recognize the need to focus on a single language.
They focus on the impact the generation methodologies of the tools can have on the final \ac{sbom}, and on the critical aspects of the \ac{sbom} itself, rather than on the impact the ecosystem can have on the generated \ac{sbom}.

\section{Discussion}
This study aims to understand the relationship between \tools and the \python software ecosystem.

We identify two root causes common to all the generation issues explored in Section~\ref{sec:results}: 
\begin{inparaenum}[(1)]
    \item The lack of standards in the \python ecosystem, and
    \item the approximation of dependency solving of \tools.
\end{inparaenum}
From them, we provide consequent recommendations.

\subsection*{Major Takeaways}
\noindent\textbf{Lack of standards for Python.}
\ac{sbom} generation is largely helped by the presence of standards.
Knowing which files must be in a project and how those files must be defined allows \tools to automatically explore the filesystem and retrieve information necessary to generate a complete and correct \ac{sbom}.
The lack of this property in Python makes \ac{sbom} generation hardly consistent with the expected standards.
The clearest example is the lack of a standard for the lockfile.
A lockfile should represent the most accurate representation of a software's dependency network;
since, however, the format varies depending on the package manager, it is problematic for \tools to use it to obtain a coherent \ac{sbom}.

\noindent\textbf{Defects in \tools.}
While the ecosystem is problematic, tools can also be blamed in various cases:
\begin{inparaenum}[\it (i)]
\item they do not consider the \texttt{pyproject.toml} file, missing dependencies in \python projects following the most recent standard;
\item they implement inaccurate version-solving techniques that affect the \sbom correctness;
\item they do not provide warnings when they cannot ensure completeness and correctness of the \sbom.
\end{inparaenum}

\subsection*{Recommendations}
Consequently to our takeaways we provide two recommendations:

\paragraph*{For the \python ecosystem}
The \python ecosystem should push for initiatives proposing standards.
Issues in \ac{sbom} generation can be largely addressed once the content of a \python project and its files is standardized.

\paragraph*{For the \tools}
\tools are required to consider the new \python standard build interface by parsing the \texttt{pyproject.toml} file.
Most \acp{sbom} will achieve completeness with this feature.

\subsection*{Additional Takeaway: Version Management}
Version management is a complex theme because the version of unpinned dependencies is solved at installation time. An \ac{sbom} created at a different time from software installation can be incorrect (e.g., the latest version of a library changed), but also incomplete (e.g., the new library version introduced a new recursive dependency). Currently, \tools seem to merely skim the issue; we think that in the future alternative solutions should be explored (e.g., deeper integration with package managers to pin all versions according to the \ac{sbom}, or associating \acp{sbom} with complete installations rather than source versions).

\section{Conclusion}
\label{sec:conclusion}
We perform a study of \ac{sbom} generation in the \python ecosystem, identifying causes of missing completeness and correctness.
We find that the lack of standards in the \python ecosystem is the main cause of inaccurate \acp{sbom}, but also that \tools do not use the \texttt{pyproject} file and do not implement proper techniques for obtaining correct dependency versions.
Based on our findings, we provide recommendations that can solve issues affecting \ac{sbom} generation for \python projects, providing better \acp{sbom} and, consequently, better transparency of the \ssc.

\printbibliography

\end{document}